\documentstyle[prl,aps,twocolumn]{revtex}

\def\prb{Phys. Rev. B}
\def\prl{Phys. Rev. Lett.}

\begin{document}
\draft
\twocolumn[
\hsize\textwidth\columnwidth\hsize\csname @twocolumnfalse\endcsname

\title
{Exact Results for 1D Kondo Lattice from Bosonization}
\author{Oron Zachar ,  S. A. Kivelson}
\address
{Dept. of Physics,
U.C.L.A.,
Los Angeles, CA  90024}
\author{V.J. Emery}
\address{
Dept. of Physics,
Brookhaven National Laboratory,
Upton, NY  11973}
\date{\today}
\maketitle

\begin{abstract}
We find a solvable limit to the problem of the 1D electron gas
interacting with a {\it lattice}
of Kondo scattering centers. In this limit, we present exact results for
the problems of incommensurate filling, commensurate filling, impurity vacancy
states, and the commensurate-incommensurate transition.
\end{abstract}
\pacs{PACS numbers:  }
]

In this paper we describe an {\it exactly}
solvable limit of a one dimensional electron gas
interacting with a {\it lattice}
of dynamic spin scattering centers.
We are interested
in this model both in its own right and for its possible applicability
to the understanding of
Kondo insulators and heavy fermion materials.  Among our most
interesting results, we find:  1)  In addition to the single impurity
Kondo scale, $\Gamma$,
there is a coherence scale, $\Delta$,
which characterizes the temperature (or frequency) below which
coherence between different impurities is established;  this scale
is also roughly equal to the gap, $\Delta_s$,
in the spin excitation spectrum which appears at low temperatures
regardless of whether the impurity lattice is commensurate, incommensurate,
or even weakly disordered.
Of course, $\Delta$ is an increasing function of impurity
concentration, $c$, and vanishes in the $c \rightarrow 0$ limit.
2)  There is a well defined crossover as
a function of $c$ from a dilute limit, in which $\Delta(c) \ll \Gamma$
to a dense limit in which $\Delta(c) \gg \Gamma$.  In the dilute limit,
single-impurity Kondo physics is apparent for temperatures
in the range $\Gamma > T > \Delta$, and coherence between impurities
sets in only for $\Delta \sim T$, while in the dense limit, coherence
and Kondo quenching of the impurity spins occur simultaneously, in
much the same way as pairing and condensation occur simultaneously
in BCS superconductors.  4)  At the crossover concentration, $c^*$,
we find that
$\Delta(c^*) \sim\Gamma$ and $c^*\sim 1/\xi_K$;
this confirms the
existence of a physical
length scale $\xi_K=v_F/\Gamma$
associated with the single impurity Kondo problem.
5)  We find that
the coherent low-temperature state is characterized by the appearance
of long-range order of a non-local order parameter, analagous to the
hidden (or topological) order found in integer spin-chains, or
the Girvin-MacDonald order parameter in the fracional quantum Hall effect.
\cite{ga}

In a discrete model, the free electron gas is described by a hopping
Hamiltonian between lattice sites with lattice constant $a$.
We consider the problem in which
the electrons interact, via Kondo scattering,
with a periodic array of localized dynamic
 spins with lattice constant $b$.
The relative impurity concentration is given by $c=a/b$.
We focus on the long-distance behavior of the electrons' correlation functions
by taking the continuum limit of the electron lattice. At first we leave the
interaction with the impurity lattice in its discrete form.
The whole system is described by the Hamiltonian
\begin{equation}
H = H_0 + H_{\parallel} + H_{\perp} .
\end{equation}
Here $H_0$ is the kinetic energy of noninteracting electrons with spin,
which, in the continuum limit is (setting $\hbar = 1$)
\begin{equation}
H_0 = -i v_F \sum_{\sigma} \int dx \left [\Psi^{\dagger}_{R,\sigma} \partial_x
\Psi_{R,\sigma} - \Psi^{\dagger}_{L,\sigma} \partial_x \Psi_{L,\sigma} \right
],
\end{equation}
where $\Psi^{\dagger}_{\lambda,\sigma}(x)$ with $\lambda = R,L$ create
respectively a right and left moving electron with z component of spin
$\sigma = \pm 1/2$ at position $x$.  $H_{\parallel} + H_{\perp}$ is the
coupling
between the electron gas and the discrete array of anisotropic Kondo impurities
 (which we refer to as ``impurity-spins''):

\begin{eqnarray}
H_{\parallel} & = & J_{\parallel} \sum_{j,\lambda} \tau_j^z \left [
\Psi^{\dagger}_{\lambda \uparrow}(R_j) \Psi_{\lambda \uparrow} (R_j) -
\Psi^{\dagger}_{\lambda \downarrow}(R_j) \Psi_{\lambda \downarrow}(R_j) \right
] \nonumber
\\ H_{\perp} & = & J_{\perp} \sum_{j,\lambda,\lambda'} \left [ \tau_j^+
\Psi^{\dagger}_{\lambda \downarrow}(R_j) \Psi_{\lambda' \uparrow}(R_j) +
{\rm H. c.} \right ] .
\end{eqnarray}
Here $\tau_j^a$ represent the dynamics of the impurity-spin
at position $R_j$.
($\tau_R^b$ satisfy canonical comutation relations
$[\tau_j^a,\tau_{j'}^b] = i \delta_{j,j'} \epsilon^{a,b,c} \tau_j^c$ ).
We will always work in the limit of large
anisotropy, $J_{\parallel} \gg J_{\perp}$.

{\bf Bosonization of the kondo lattice: }
We bosonize the Hamiltonian using the standard one dimensional
relation between Bose and Fermi fields \cite{Boso1D}:
\begin{equation}
\Psi^{\dagger}_{\lambda,\sigma}(x)= \frac {1} {\sqrt {2\pi a}}
exp\{i\Phi_{\lambda,\sigma}(x) \}
\end{equation}
where
$\Phi_{\lambda,\sigma} =
\sqrt{\pi} [ \theta_{\sigma} (x) \pm \phi_{\sigma}(x) ]
\pm k_F x$
with ``+'' and ``-'' corresponding to $\lambda=Left$ and $Right$
respectively, $\theta_{\sigma}(x)= \int_{- \infty}^x dx' \Pi_{\sigma}(x')$,
$\phi_{\sigma}(x)$ and $\Pi_{\sigma}(x)$ are canonically conjugate Bose
fields, so that $[\phi_{\sigma}(x),\Pi_{\sigma'}(x')] = i \delta(x-x')$.
To take advantage of the separation between
charge and spin which occurs in one dimension, we express the Hamiltonian
in terms of a spin field, $\phi_s(x) = [\phi_{\uparrow}-\phi_{\downarrow}]/
\sqrt{2}$, and a charge field, $\phi_c(x) =
[\phi_{\uparrow}+\phi_{\downarrow}]/
\sqrt{2}$ and correspondingly defined momenta,
$\Pi_s$ and  $\Pi_c$.
We introduce distinct coupling constants
which we label with the superscripts ``f'' and ``b''
for the forward and back-scattering terms in the
Hamiltonian.
This distinction
can be associated with the form factor
of a spatially extended impurity. The result is easily seen to be:

\begin{equation}
H_0 = \frac {v_F} 2 \int dx \left\{ [ \Pi_{c}^2 +(\partial_r\phi_c)^2  ]
+ [ \Pi_{s}^2 +(\partial_r\phi_s)^2  ] \right \}
\end{equation}
\begin{eqnarray*}
H_{\parallel} & = & \ J_{\parallel}^f \sqrt{\frac 2 {\pi}} \sum_j\tau_j^z
\partial_r\phi_s(r_j)       \nonumber
\\ & + &\frac {2 J_{\parallel}^b} {a\pi} \sum_j \tau_j^z
\sin[\sqrt{2\pi}\phi_s(j)] \sin[\sqrt{2\pi}\phi_c(j) + 2 k_F R_j] , \nonumber
\\ H_{\perp} & = & \sum_j \tau_j^+  e^{-i\sqrt{2\pi}\theta_s(j)}
 \{ \frac {J_{\perp}^f} {\pi a}  \cos[\sqrt{2\pi}\phi_s(j)]
\\ & + &
\frac {J_{\perp}^b} {\pi a}  \cos[\sqrt{2\pi}\phi_c(j) + 2 k_F R_j]
 \}     +   H.c.  .
\end{eqnarray*}
We eliminate the explicit dependence on the $\theta_s$ field by making a
unitary transformation \cite{EK1,EKlatt}
\begin{equation}
U = \exp{ [-i\sqrt{2 \pi}\sum_j \tau_j^z \theta_s(R_j) ]}.
\end{equation}
The resulting transformed Hamiltonian is $ \tilde H = U^{\dagger} H  U $

\begin{eqnarray}
\tilde H_0 & = & H_0^s + H_0^c
 + \Delta J_{\parallel}^f a \sqrt{\frac 2 {\pi}}
\sum_R\tau_R^z \partial_x\phi_s(x_j)  ,             \nonumber
\end{eqnarray}

\begin{eqnarray}
\tilde H_{\parallel}= \frac {2 J_{\parallel}^b} {a\pi} \sum_j
&& \tilde \tau_j^z (-1)^j
\nonumber
\\ & \times & \sin[\sqrt{2\pi}\phi_s(j)] \sin[\sqrt{2\pi}\phi_c(j) + 2 k_F R_j]
\nonumber
\\ \tilde H_{\perp} = \frac {J_{\perp}^f} {\pi a} \sum_j
&& \tilde \tau_j^x (-1)^j  \cos[\sqrt{2\pi}\phi_s(j)]
\nonumber
\\  + \frac {J_{\perp}^b} {\pi a} \sum_j
&& \tilde \tau_j^x \cos[\sqrt{2\pi}\phi_c(j) + 2 k_F R_j]   ,
\end{eqnarray}
where $\tilde \tau_j^x \equiv U \tau_j^x U^+$, and
$\Delta J_{\parallel}=J_{\parallel}^f -\pi v_F$.
We see that the result of the unitary transformation is to adsorb a spin degree
of freedom (the bosonic phase field $\theta_s$) into the impurity operator
$\tau^+$ leaving behind a phase shift $\pi$ of the number field
$\sqrt{2\pi}\phi_s$ between impurity sites (giving rise to the staggered
coefficient $(-1)^j$).
We interpret this as a description of the Kondo resonance in
the bosonization language.
Therefore we can interpret the unitary transformation as a mapping from
a free-field to a strong coupling basis. In the new basis the Hamiltonian is
so simple that much of the physics can be read directly from it.

{\bf Incommensurate phase: }
The solvable limit \cite{EK1,EKlatt}
$J_{\parallel}^f = v_F \pi$ (i.e. $\Delta J_{\parallel} = 0$)
is analogous to the Toulouse limit of the single impurity Kondo problem.
For incommensurate filling, $2 k_F R_j \neq n \pi$, we make the approximation
of neglecting all terms in the Hamiltonian
which depend on $2 k_F R_j$. (We will return to this point, below,
when we consider the commensurate-incommensurate transition).
In this limit,

\begin{eqnarray}
\tilde H_{inc} = & & H_0^s +
\frac {J_{\perp}^f} {\pi a} \sum_j \tilde \tau_j^x (-1)^j
\cos[\sqrt {2 \pi} \phi_s(j)] + H_0^c .
\end{eqnarray}
$\{ \tilde \tau_j^x \}$
commute with the Hamiltonian, so we can regard them as
c-numbers $( \tilde \tau_j^x = \pm {\frac 1 2} )$,
labeling eigenstates of the transformed Hamiltonian with static configuration
of $\tilde \tau_j^x$.
Also, the charge and spin parts of the Hamiltonian decouple.
The charge excitations, $\phi_c$, are just the gapless free field states of
$H_0^c$.
The spin Hamiltonian has a discrete sine-Gordon form. Due to the $(-1)^j$
factors, the ground-state is one in which the transformed impurity
spins $\tilde \tau_j^x $ are antiferromagnetically ordered and the
$\phi_s$ spectrum is gapped, $E_s(k)=\pm \sqrt{(v_F k)^2 + \Delta_s^2}$.
With respect to the ground state of this reduced Hamiltonian,
small perturbations in
$\Delta J_{\parallel}$  and $J_{\parallel}^b$ are irrelevant.
Therefore, results obtained in this limit are generic for at least a finite
region of parameter space.

Physically there are two important energy scales: The single impurity Kondo
resonance
energy scale $\Gamma$, (which in the Toulouse limit is
$\Gamma = J_{\perp}^2 / \pi v_F a$), and the inter-impurity coherence scale
$\Delta$ which is set by the spin gap, $\Delta_s$, calculated below.
The interplay
between the two energy scales is sensitive to the impurity concentration.

{\it The Dense Limit - } Associated with the spin gap $\Delta_s$ in
the spectrum there is a corresponding fermionic correlation length
$\xi_s = v_F/\Delta_s$. When $\xi_s$ is much bigger than the characteristic
distance between impurity sites, $b = a/c \ll \xi_s$, the discrete character
of the imurity array can be ignored. Therefore, in this limit (high density),
we can also take the continuum limit with respect to the impurity lattice.
The spin Hamintonian now takes the regular sine-Gordon form
\begin{equation}
\tilde H_s = H_0^s + \frac {c J_{\perp}^f} {2 \pi a^2} \int dx
\cos[\beta \phi_s(x)]  ,
\end{equation}
with $\beta = \sqrt {2 \pi}$. The resulting gap is \cite{Boso1D}
\begin{equation}
\Delta_s \sim
\frac {v_F} a \left[c J_{\perp}^f/v_F\right]^{2/3}.
\end{equation}
Similarly, there is an energy gap $\Delta \sim \Delta_s$
for creation of kinks in the transformed impurity spin order.

Thus, the system is dense if $c$ exceeds the critical density $c_1$,
at which $\xi_s (c_1) = a/c_1$;  consequently, in this limit
$\Delta_s > \Gamma$. Since
$\Delta_s$ also determines the temperature scale below
which coherence sets in,
this is analogous to the BCS limit of superconductivity, in that local
coherence
between the impurity spins and the conduction electrons, and longer range
coherence between the impurity spins occur at the same temperature.

{\it Dilute limit -}
When the distance between impurities is large, $a/c \gg a$,
the discrete nature of the impurity array cannot be ignored.
For small enough impurity concentrations $c<c_2$,
we can compute \cite{EKlatt}
the leading order dependence of $\Delta$ in powers of the distance between
impurities;
$\Delta = \left( {\pi v_F c} /{4 a}\right)
\left[ 1 + {\cal O}(c v_F/J_{\perp})
\right]$.
The
consistency of the dilute limit is determined by the simultaneous conditions,
$c v_F/J_{\perp} \ll 1$ and $\Delta \ll \Gamma$,
of which the latter is the more restrictive by a factor $v_F/J_{\perp}$.
{}From this condition we get
$c_2\sim \left(  {J_{\perp}} /{\pi v_F} \right) ^2$.
Note that there is actually only one characteristic concentration,
$c_1 \sim c_2 \sim c^*$.
In the low temperature coherent regime, $T \ll \Delta$,
the behavior of
both dilute and dense systems are qualitatively
indistinguishable;  for simplicity, we will henceforth deal with
systems in the dense limit.

{\it Order parameters -}
The most interesting physical quantities
the correlation functions of  the
various possible order parameter fields.
(Since this is a one dimensional system, we will never see a broken continuous
symmetry, but we can interpret the emergence of fairly long range coherence in
the fluctuations of a dominant collective field as the one dimensional analog
of the phase transition to a broken symmetry state that we would expect in
higher dimensions.)
In the case of noninteracting electrons, density-density
correlation functions decay as $1/x^2$. Therefore, any of the various
possible order parameters whose correlation functions
$C_i(x,x') =<O_i(x)O_i(x')>$
decay like $x^{- \alpha_i}$ with $\alpha_i < 2$,
can be considered as having enhanced long range coherence.
The enhanced order parameters are: $2 k_F$ charge density wave
$O_{CDW}=[\Psi_{L,\uparrow}^{\dagger} \Psi_{R,\uparrow} +
\Psi_{L,\downarrow}^{\dagger} \Psi_{R,\downarrow}]$,
composite spin density wave
$O_{cSDW}=[\Psi_{L,\downarrow}^{\dagger} \Psi_{R,\uparrow} \tau_{j(x)}^+]$
(where $j=j(x)$ is the j-th site nearest to position x),
singlet pairing
$O_{SP}=[\Psi_{L,\uparrow}^{\dagger} \Psi_{R,\downarrow}^{\dagger}]$, and
composite triplet pairing
$O_{cTP}=[\Psi_{L,\downarrow}^{\dagger} \Psi_{R,\downarrow}^{\dagger}
\tau_{j(x)}^+]$. Each exponent $\alpha_i$ has contributions from the
spin and charge
correlation function exponents $K^*_s$ and $K_c$ as follows:
$\alpha_{CDW} = K_c + K^*_s$,
$\alpha_{SP} = 1/K_c + K^*_s$,
$\alpha_{cSDW} = K_c$, and
$\alpha_{cTP} = 1/K_c$.
Since the spin field is massive,
the spin-fluctuations are frozen out,  which
results in
$K^*_s = 0$ in these correlation functions, and causing exponential
decay of others.   Moreover,
deep in the incommensurate phase the charge fields are effectively free, so
$K_c = 1$. Note that the
unitary transformation has
eliminated the spin field dependence of the composite
order parameters.

All the above order parameters have staggered order
$C_i (x,x') \sim (-1)^{(j-j')} |x-x'|^{- \alpha_i}$.
The impurity correlation function $<\tau_j^x  \tau_{j'}^x>$ decays
exponentialy.
 However, the transformed impurity spin correlation function,
$<\tilde \tau_j^x \tilde \tau_{j'}^x> \sim const.$ , exhibits long-range order
at $T=0$;  this
is the  non-local order parameter which characterizes the coherent state.
\cite{ga}

The composite triplet pairing is especially interesting.
Order parameters of this type were previously found in the solution of the
single impurity 2-channel Kondo problem \cite{EK1}, but not in
the single channel Kondo problem. In going to the incommensurate limit we have
neglected all the  backscattering terms. If we consider the
Kondo interaction without backscattering then it can be written as

\begin{eqnarray}
H_{impur} = & & J^f \sum_{j,i} {\boldmath {\tau}} _R^z \cdot \left [
\Psi^{\dagger}_{i \alpha}(R_j) {\boldmath {\sigma}} _{\alpha \beta}
\Psi_{i \beta}(R_j) \right ]
\end{eqnarray}
where i=R,L. The impurity scatters the right and left going electrons as if
they were two independent channels. Of course, for the single impurity problem,
$J^b$ would lift this artificial channel degeneracy, and hence is a relevant
pertubation. Even if $J^f \ll J^b$, the single impurity is ultimately governed
by the single-channel fixed point. However, for the array, the existence of the
spin gap stabilizes the two-channel behavior of the model with $J^b=0$,
rendering small $J^b$ an irrelevant perturbation.
We would like to note that a composite triplet pairing order parameter of a
different kind was found in a mean field theory of isotropic the 3D Kondo
lattice \cite{cmt}.
The relation between the two results is not clear at the present.

{\bf Commensurate filling: }
In the case where $k_F$ is commensurate with the impurity lattice,  the back
scattering terms cannot be ignored.
In particular we will now discuss the case
$2 k_F R_j = j \pi$ (usually called half-filling). The transformed Hamiltonian
is then

\begin{eqnarray}
\tilde H & = & H_0^s +  \frac {1} {\pi a}
 \sum_j \tilde \tau_j^x (-1)^j J_{\perp}^f \cos[\sqrt{2\pi}\phi_s(j)] \nonumber
\\ & + & H_0^c + \frac {1} {\pi a}
 \sum_j \tilde \tau_j^x (-1)^j J_{\perp}^b \cos[\sqrt{2\pi}\phi_c(j)] \nonumber
\\ & + & \ \frac {2 J_{\parallel} ^b } {a \pi} \sum_j \tau_j^z
\sin[\sqrt{2\pi}\phi_s(j)] \sin[\sqrt{2\pi}\phi_c(j)]
\end{eqnarray}
In the limit $J_{\parallel}^b =0$, $\tilde \tau_j^x$ are good quantum numbers.
The spin and charge decouple into two independent sine-Gordon Hamiltonians,
 which are minimized when the $\phi$ fields take the values $2\pi n$.
Consequently, both spin and charge excitation spectra are gapped, and
$\tilde \tau_j^x$ are ordered antiferromagnetically. The quantum numbers of the
excitations are easily determined in the bosonization formalism.  Solitons in
the $\phi_s$ field are $\sqrt {2 \pi}$ phase slips, amounting to a state with
spin $S=1$ and charge $Q=0$.
By the same analysis, a $\phi_c$ soliton carries $S=0$ and $Q=2e$.
A kink (or domain wall) in the $\tilde \tau_j^x$ order induces
compensating $\frac 1 2 \sqrt {2 \pi}$ phase slips in
$\phi_c$ and $\phi_s$, resulting in a
bound state with charge Q=1 and spin S=1/2.
As could be expected for commensurate filling, the only order  parameter which
does not decay exponentially is the $2 k_F$ CDW order;
  $O_{CDW}=[\Psi_{L\uparrow}^{\dagger} \Psi_{R\uparrow} +
  \Psi_{L\downarrow}^{\dagger} \Psi_{R\downarrow}]$.
At long distances $<O_{CDW}(x) O_{CDW}(x')> \sim$ constant,
since all fields
 $\phi_s, \phi_c, \tilde \tau_j^x$ are gapped. Deviations
$J_{\parallel}^b \neq 0$ can be treated perturbatively giving rise to short
range hopping dynamics of $\tilde \tau_j^x$ kinks.

Note that there is a second possible stable phase when
$J_{\parallel}^b \neq 0$ in which
$\tau^z$ orders and the $\phi$ fields take the values $2\pi n + 1/2$.
This second stable phase is analagous to the
dimerized commensurate phase of a half-filled Peierls insulator,\cite{ssh}
such as polyacetylene; its implications will be explored
elsewhere.

{\bf Vacancy states: }
What happens when we introduce imperfections by
deleting a single impurity from the periodic array?
The perfect array described by the Hamiltonian in equation (12) had a staggered
factor $(-1)^j$ in front of both $\cos[\sqrt{2\pi}\phi_s(j)]$ and
$\cos[\sqrt{2\pi}\phi_c(j)]$ terms, but the origin of this factor is different
for each term.
The $(-1)^j$ factor in the $\phi_s$ part of the Hamiltonian was a result of the
unitary transformation; it ``counts'' impurities, but is insensitive to the
distance between impurities. The $(-1)^j$ in the $\phi_c$ part came from the
$2 k_F R_j$ phase in
$\cos[\sqrt{2\pi}\phi_c(j) + 2 k_F R_j]$, and therefore explicitly depends
upon the distance between Kondo spins.  Across a Kondo vacancy there
will be a conflict between the impurity $\tilde \tau_x^j$ order induced by the
spin fields and the charge field. This can be amended by either twisting the
$\phi_s$ or the $\phi_c$ fields across the vacancy site.
Therefore, associated with a vacancy there could be two alternative, mutually
exclusive, bound states: Either a local $\phi_c$ soliton of Q=1 and S=0, or
a local $\phi_s$ soliton with Q=0 and S=1/2.  This is analgous to the
solitons in polyacetylene.\cite{ssh}

{\bf Commensurate-Incommensurate Transition:}
For incommensurate wave number $k_F$ we can define incommensurability q such
that $(k_F + q) R_j = j \pi$. Making a simple change of variables
$\phi_c $  $\rightarrow$  $ \phi_c - 2 qx/\sqrt{2 \pi}$,
the transformed incommensurate charge part of the Hamiltonian can be rewritten
in the form

\begin{eqnarray*}
\frac {\tilde H^c} {v_F} & = & \frac {1} 2 \int dx \left [
\frac {\delta ^4} {u_p^2} \Pi_{c}^2 +(\partial_x\phi_c - \delta )^2
\right ]   \nonumber
 + \int dx \frac {\delta ^2} {h^2} \cos[\beta \phi_c] ,
\end{eqnarray*}
where $\beta = \sqrt{2\pi}$, $\delta = \frac {2 q} {\beta}$, $u_p = \delta ^2$,
$\frac {h} {\delta} = \sqrt {\pi a^2 v_F /J_{\perp}^b}$.
Classically, this Hamiltonian is known to have an incommensurate periodic
soliton-lattice solution \cite{Fetter}
above a critical incommensurability $q_c$

\begin{equation}
q_c = \frac {2 \beta} {\pi a} \sqrt {\frac {J_{\perp}^b} {\pi v_F}} ,
\end{equation}
and to remain commensurate for $q < q_c$.
Thus, for $q > q_c$ there is a gapless
Goldstone mode (acoustic vibrations of the soliton lattice).
Note that the acoustic velocity $u_p$ is not equal to $v_F$;
it gradually increases with $\delta$.
The solitons have a width
$L_s = \frac {a} {\beta} \sqrt {\pi v_F /J_{\perp}}$, so when
the distance between solitons become comparable with soliton width, there will
be a ``melting'' crossover to the fully incommensurate state discussed before.
In our model this crossover will happen at incommensurability
$q' \sim \pi^2 q_c$, so there is a
significant range in which the soliton lattice exists.

The gapless Goldstone modes, associated with the broken symmetry
soliton lattice ground state, are not trivially related to the free
$\phi_c$ fields. Therefore, we can expect
correlation functions of operators like $e^{i \beta \phi_c (x)}$ (i.e. the
order parameters) to have anomalous exponents.
Indeed, following the renormalization group arguments of
Schulz \cite{IncommExponents},
an incommensurate umklapp term renormalizes $K_c$,
the charge field contribution
to the exponents of correlation functions. The renormalized exponent,
$1/2 < K^*_c < 1$, tends to 1/2 for $q \rightarrow q_c$ and to 1 for $q
\gg q'$.

We now address the question of deviation from perfectly periodic impurity
array.
Since the spin part of the Hamiltonian is completely insensitive to the
distance between impurities, the spin excitations will be gapped even in the
case of a randomly spaced impurity array (i.e. disordered array).
The charge field, on the other hand, is affected by the disorder.
Following the method of Giamarchi and Schulz \cite{disorder1D}
we conclude that a disordered 1D Kondo array will be insulating.
Therefore, the 1D disordered Kondo array will be a gapless charge
insulator with a spin gap.

{\bf Acknowlegments:}  This work was supported in part by
the National Science Foundation under
grants number PHY94-07194 at the Institute for Theoretical Physics (SAK)
and DMR93-12606 at UCLA (SAK and OZ) and by the Department
of Energy under Contract No. DE-AC02-76CH00016 (VJE).

\end{document}